\documentclass[]{pasj01}
\draft

\Received{}
\Accepted{}
 
\begin{document} 

\title{ 
Mass flows and their behaviors in the SS433 -- W50 system }

\author{Hajime \textsc{Inoue}\altaffilmark{1}}%
\altaffiltext{1}{Institute of Space and Astronautical Science, Japan Aerospace Exploration Agency, 3-1-1 Yoshinodai, Chuo-ku, Sagamihara, Kanagawa 252-5210, Japan}
\email{inoue-ha@msc.biglobe.ne.jp}



\maketitle

\begin{abstract}
We propose the scenario to interpret the overall observational features of the SS433 -- W50 system.
The most unique features of SS433 are the presence of the precessing, mildly relativistic jets and the obscuration of the central engine, which are considered to be due to a supercritical accretion on to the central compact object. 
The jets are likely to be ejected from the innermost region of the accretion flow.
The concept of the accretion ring (Inoue 2021, PASJ, 73,795) is applied to the outer boundary of the accretion flow and the ring is supposed to have a precession.
The accretion ring is expected to extend a two-layer outflow of a thin excretion disk and a thick excretion flow, as well as the accretion flow.
The thin excretion disk is discussed to eventually form the optically thick excretion belt along the Roche lobe around the compact object, contributing to the obscuration of the central engine.
The thick excretion flow is likely to turn to the supersonic wind (disk wind) with the terminal velocity of $\sim 10^{8}$ cm s$^{-1}$ and to collide with the SNR matter at the distance of $\sim 10^{18}$ cm.
The interactions of the jets with the disk wind are considered to cause the features of the jets observed at the distances of 10$^{14} \sim 10^{15}$ cm and $\sim 10^{17}$ cm.
Finally, it is discussed that the jets are braked by the SNR matter at the distance of $\sim$10 pc and the momentum carried by the jet is transferred to the SNR matter shoved by the jet.
The SNR matter pushed to the inside of the precession cone is expected to gather along the cone axis and to form the elongated structures in the east and west directions from the main W50 structure.

\end{abstract}

\section{Introduction}\label{Introduction}
SS433 is well known to steadily exhibit the precessing, mildly relativistic jets.
The motion was discovered in the observations of optical emission lines and soon found to be nicely explained by the kinematic model, in which matter is ejected in two opposing jets that are collimated and oppositely aligned to within a few degrees (see Margon 1984 for the early review).  The ejection velocity proves to be 0.26 $c$ ($c$ : the velocity of light), the axis of the jets rotates with a period of 162.5 days, the central axis of the rotation cone is inclined by 79$^{\circ}$ from the line of sight, and the half-angle of the cone is 20$^{\circ}$.

Further observations revealed presence of two other periodicities in the SS433 system than the preccessing period: the binary period of 13.1 days (Crampton et al. 1980; Cherepashchuk 1981) and the nodding period of 6.3 days (Katz et al. 1982).

From observational properties periodically varying with the binary period, a number of studies have been done to determine whether the compact object in the SS433 system is a neutron star or a black hole (see subsections of 7.2 and 7.3 of Fabrika 2004 for the brief review as of early 2000; Cherepashchuk et al. 2018 and references therein for the recent progresses).
Although recent studies favor the black hole hypothesis, it is not conclusive yet.

The bolometric luminosity of SS433 is of the order of 10$^{40}$ erg s$^{-1}$ (see subsection 6.6 of Fabrika 2004 and references therein), indicating that a supercritical accretion on to a compact object takes place in the system.
The observed spectrum is, however, approximated by a blackbody spectrum with the temperature of several 10$^{4}$ K and the blackbody radius is as large as a few times 10$^{12}$ cm.
This emission is understood to be a result of reprocessing the direct emission from the central engine by an obscuring matter on the line of sight.

According to the current observational study of binary X-ray sources (see e.g. Inoue 2022a for the review on X-ray observations of accretion disks), the emission from an accretion flow on to a compact object is mainly observed in the soft X-ray ($\sim$ 1 - 10 keV) band.
The observed soft X-ray luminosity from SS433 is, however, as low as $\sim 10^{36}$ erg s$^{-1}$ and is, furthermore, realized to come from the jets ejected from the central engine.
It is considered that the soft X-rays from the central engine is completely obscured and 
that only the hard X-rays above $\sim$20 keV with the luminosity of several times 10$^{35}$ erg s$^{-1}$ is observed from an extended hot corona on an accretion disk (Cherepashchuk et al. 2009; 2013).

Several emission lines are detected in the soft X-ray band and well interpreted as results of thermal emission from adiabatically and radiatively cooling conical jets with distance of the order of 10$^{12}$ cm from the center (Kotani et al. 1996). 

With the detailed jet model, Brinkmann and Kawai (2000) estimated the kinetic luminosity of the jets $\sim 6 \times 10^{39}$ erg s$^{-1}$.

The X-ray jets end at a distance of $2 \sim 3 \times 10^{13}$ cm and the optical jets replace the X-ray jets on the further side.
The optical emission from the jets often exhibits time variations on a time scale of days, which are interpreted by appearances of cool clouds and their modulations in the jets as a result of thermal instability induced in the jet where the temperature decrease to $\sim$0.1 keV.
The active region of the optical lines from the clouds begins $\sim 10^{14}$ cm, has the activity-maximum around several times 10$^{14}$ cm and ends at $\sim 10^{15}$ cm.
For the review on the properties of the optical jets, see sections 2 of Fabrika (2004).

The lifetimes of the moving clouds emitting optical lines are observed to be several days, while the cooling time is estimated to be much shorter that them.
Thus, the matter must be continuously reheated.
The interaction of the jets with the accretion-disk wind is currently thought to be the most likely origin for the heating (see subsection 5.3 of Fabrika 2004 and references therein).

Radio blobs are resolved in the optical-jet region with the VLBI observations. 
The radio strength appreciably weakens as they moves outward but grows again as they pass through the ``radio brightening zone" with the distance of $\sim 4 \times 10^{15}$ cm from the center.  
The radio blobs rapidly fade beyond the brightening zone.
For the review on the radio properties of the jets, see sections 3 of Fabrika (2004).

The radio interferometric observations also reveal the existence of matter in the plane perpendicular to the jets (Paragi et al. 1999; Blundell et al. 2001).
The structure extends to $3 \sim 4 \times 10^{15}$ cm on both sides of the central source.
Outward proper motion of a radio component is detected and the speed is estimated to be $\sim 10^{8}$ cm s$^{-1}$.

The brightness of the radio jets becomes the maximum at a distance of $\sim 10^{15}$ cm and gradually decreases to $\sim 10^{17}$ cm, beyond which the jets are not visible until $\sim 10^{20}$ cm.
The X-ray imaging observation with Chandra, however, indicates that the jets are reheated so hot as to emit X-rays at the distance of $\sim 10^{17}$ cm (Migliari et al. 2002).

The jets terminate at a distance of several 10 pc from the center,  in the elongated structures in the east and west directions of the radio nebula W50.
W50 has a shell like structure with radius of $\sim$ 40 pc except the elongated regions and is thought to be a supernova remnant (see Margon 1984).
Using high-resolution radio imaging observations spanning a 12-year period, the proper motions of the radio filaments across the W50 nebula were analyzed and the upper limit of $\sim 1 \times 10^{9}$ cm s$^{-1}$ was obtained for them (Goodall et al. 2011). 

The elongated structures extending to a distance of $\sim$ 100 pc are considered to be formed as a result of interactions between the jets and the interstellar matter.
The eastern and western optical filaments lie inside the projection of the jet preccession cone and are perpendicular to the jet.
Extended X-ray jets are also detected with fringing the optical filaments (Watson et al. 1983).

In this paper, we try to interpret the observational properties of the SS433 - W50 system as briefly reviewed above as widely as possible by a model situation.
The model situation is introduced in section 2, interpretations of the observations are presented in section 3 and the summary is given in section 4.

\section{Model situation}
We suppose the following situation.

A binary system composing of a normal star and a compact object and having the orbital period of 13.1 days stays at the center of a supernova remnant having a spherical shell with a radius of $\sim$40 pc.

Although the the masses of the two stars have not precisely been determined from observations yet, we assume the compact object as a black hole with $\sim10\; M_{\odot}$ and adopt $\sim20\; M_{\odot}$ as the typical mass of the normal star here.  
These assumptions have no large effect on the present study.

The separation of the two components, $a$, in the binary with the orbital period of 13.1 days is estimated as
\begin{equation}
a = 4.4 \times 10^{12} (1+q)^{1/3} \left(\frac{M_{\rm n}}{20\ M_{\odot}}\right)^{1/3} \rm{cm},
\label{eqn:a}
\end{equation}
where $q$ is the ratio of the compact star mass, $M_{\rm x}$, to the normal star mass, $M_{\rm n}$.
Considering $M_{\rm n} \sim 20 M_{\odot}$ and $q < 1$, the separation is several times 10$^{12}$ cm.

In the binary, matter flows from the normal star through the L1 point to the compact object and forms an accretion ring at the Keplerian circular orbit around the compact object determined by the specific angular momentum of the inflowing matter.
According to the approximations done in subsections 4.4 and 4.5 of Frank, King and Raine (2002), the radius of the accretion ring, $R_{0}$, is roughly given as
\begin{equation}
\frac{R_{0}}{a} \simeq \frac{1+q}{q} (0.5 + 0.227 \log q)^{4}, 
\label{eqn:R_0}
\end{equation}
and is estimated to be $\sim$0.10 for $q \sim 0.5$.
On the other hand, the radius of the Roche lobe on the side of the compact star, $R_{1}$, is approximated as
\begin{equation}
\frac{R_{1}}{a} \simeq \frac{0.49 q^{2/3}}{0.6q^{2/3}+\ln (1+q^{1/3})}
\label{eqn:R_1}
\end{equation}
(Eggleton 1983), and is 0.32 for $q$ = 0.5.
For $a \sim$ several times 10$^{12}$ cm, $R_{0}$ and $R_{1}$ are considered to be several times 10$^{11}$ cm and $\sim 10^{12}$ cm respectively.

\begin{figure}
 \begin{center}
  \includegraphics[width=14cm]{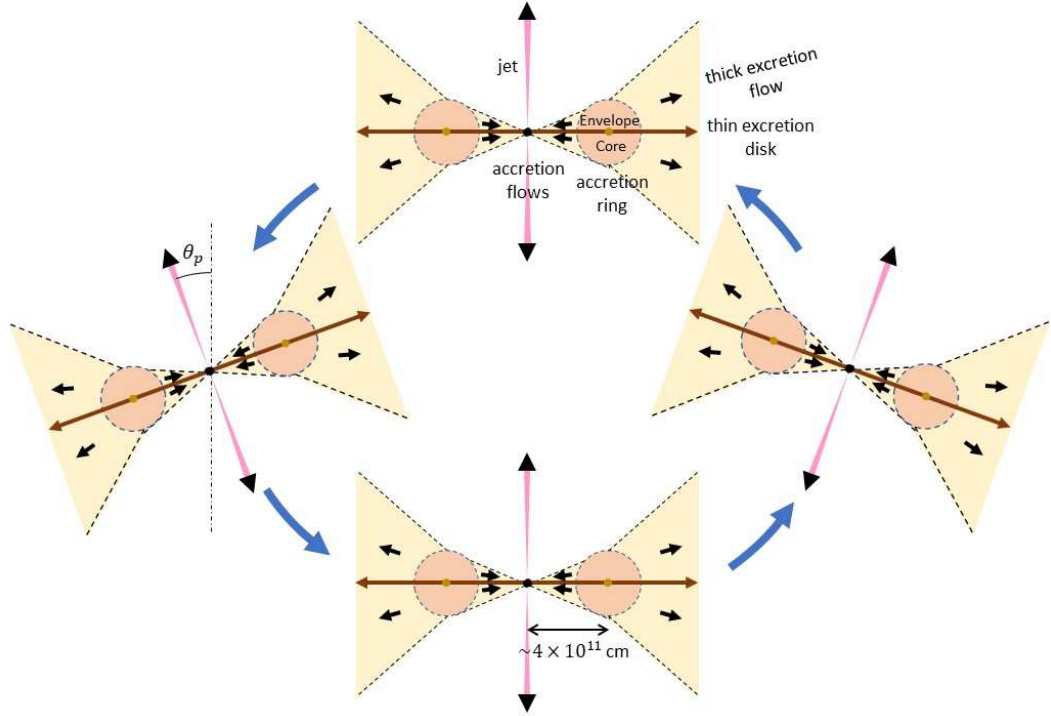}  
 \end{center}
\caption{The schematic cross section of the central part of the system considered in the present scenario.  The four diagrams indicate the situation in which the central part precesses around the angular momentum axis of the accretion ring. }\label{fig:AccretionRing}
\end{figure}

According to the recent study by Inoue (2021a), it is supposed that the accretion ring consists of a geometrically thick envelope and a geometrically thin core, and that angular momenta are transfered from the inner side facing to the compact object to the opposite side respectively in the envelope and the core. As a result, a thick accretion flow and a thick excretion flow extend from the envelope, and a thin accretion disk and a thin excretion disk do from the core. 
The schematic configuration around the accretion ring in SS433 is shown in figure \ref{fig:AccretionRing}.

Let us define $\dot{M}_{0}$, $\dot{M}_{\rm ac,\; f}$, $\dot{M}_{\rm ex,\; f}$, $\dot{M}_{\rm ac,\; d}$, and $\dot{M}_{\rm ex,\; d}$ as the mass inflow rate to the accretion ring from the normal star, and the mass flow rates of the thick accretion flow, the thick excretion flow, the thin accretion disk and the thin excretion disk, respectively.
Then, we assume that
\begin{equation}
\dot{M}_{\rm ac,\; f} = \dot{M}_{\rm ex,\; f} = \frac{F\; \dot{M}_{0}}{2},
\label{eqn:Mdot_f}
\end{equation}
and
\begin{equation}
\dot{M}_{\rm ac,\; d} = \dot{M}_{\rm ex,\; d} = \frac{(1-F)\; \dot{M}_{0}}{2}.
\label{eqn:Mdot_d}
\end{equation}
Here, $F$ is the fraction of the total flow rate to the thick accretion and excretion flows in the total inflow rate, and is supposed to be several to 10\% when the inflow rate is as large as or larger than the Eddington rate (Inoue 2021a).
The equalities of the mass flow rate between the accretion flows and the excretion flows are derived from the study by Inoue (2021a; see also the appendix in Inoue 2021b).

For the accretion flow on to the central compact object, a situation is expected to happen in which a geometrically thin standard disk (Shakura \& Sunaev 1973) is sandwiched by a geometrically thick and optically thin ADAF (Narayan \& Yi 1994).
The thin standard disk is considered to turn to an optically thick ADAF called as the slim disk (Abramovicz et al. 1988) in the inner region because of the supercritical accretion rate. 

Jets are ejected from the innermost region of the slim disk in two opposite directions perpendicular to the equatorial plane of the disk.
The jet velocity, $v_{\rm j}$, is assumed to be 0.26$\; c$ as observed.

The flow rate of the specific angular momentum on to the black hole is $\sim \sqrt{r_{\rm in}GM}$ for the innermost radius of the accretion flow, $r_{\rm in}$, and is negligibly small compared with the inflow rate to the accretion ring, $j_{0}$, given as
\begin{equation}
j_{0}= \sqrt{R_{0}GM},
\label{eqn:j_0}
\end{equation}
since $r_{\rm in} \ll R_{0}$.
Thus, almost all the angular momentum of the accreted matter should be transfered to the excreted matter across the boundary layer in the accretion ring, irrespectively of whether the flow is thick or thin.
As the result, the specific angular momentum carried by the excreted matter, $j_{\rm ex}$, should be twice as large as the intrinsic inflow rate, $j_{0}$, written as
\begin{equation}
j_{\rm ex} \simeq 2\; j_{0}.
\label{eqn:j_ex-j_0}
\end{equation} 

The angular momentum transfer is done with stress due to turbulent viscosity in the boundary layer and the total torque given to the innermost side of the excretion flow is $\dot{M}_{\rm ac}\; j_{0}$.
Hence, the energy transfer rate in association with the angular momentum transfer to the excretion flow should be $\dot{M}_{\rm ac} j_{0} \Omega_{0}$, where $\Omega_{0}$ is the angular velocity of the inflowing matter at $R_{0}$.
Considering $\dot{M}_{\rm ex} = \dot{M}_{\rm ac}$, $j_{0}=\sqrt{R_{0}GM}$ and $\Omega_{0}=\sqrt{GM/R_{0}^{3}}$, the specific energy given to the excretion flow, $\varepsilon_{\rm out}$, is estimated as
\begin{equation}
\varepsilon_{\rm out} = j_{0}\; \Omega_{0} \simeq \frac{GM}{R_{0}}.
\label{eqn:epsilon_out}
\end{equation}

\begin{figure}
 \begin{center}
  \includegraphics[width=10cm]{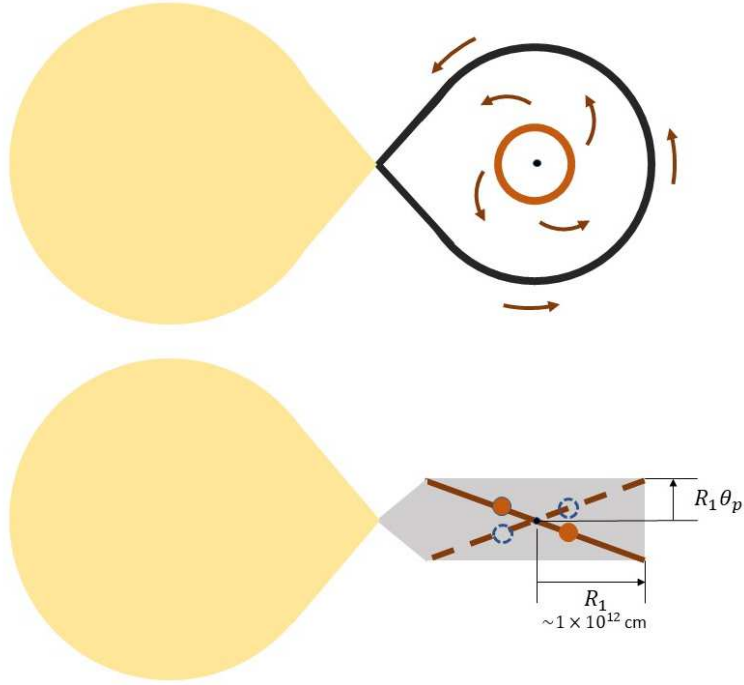}  
 \end{center}
\caption{The schematic view of the excretion belt from top (upper) and on the meridian cross section (bottom).}\label{fig:ExcretionBelt}
\end{figure}

In case of the thin excretion disk, the matter slowly outflows by adjusting the centrifugal force to balance with the gravitational force and radiating the surplus energy via the blackbody emission.
The outflow of the excretion disk should terminates at the Keplerian circular orbit, $r_{\rm ex}$, determined by the specific angular momentum $j_{\rm ex}$ given in equation (\ref{eqn:j_ex-j_0}), which is calculated as
\begin{equation}
r_{\rm ex} = \frac{j_{\rm ex}^{2}}{GM} \simeq 4 \frac{j_{0}^{2}}{GM} = 4 R_{0}.
\label{eqn:r_ex}
\end{equation}
Since the Roche lobe radius around the compact star, $R_{1}$, is calculated to be $\sim 3 R_{0}$ for $q \sim 0.5$ from equation (\ref{eqn:R_0}) and (\ref{eqn:R_1}), 
this radius, $r_{\rm ex}$, is slightly larger than the Roche lobe radius around the compact star.
Thus, the geometrically thin excretion disk would, in practice, significantly suffer the tidal effect from the normal star and could form a belt-like structure along the Roche lobe, eventually merging with the normal star atmosphere, as schematically depicted in figure \ref{fig:ExcretionBelt}.
We call this structure ``excretion belt" hereafter.

In case of the thick excretion flow, the flow is considered to be advection dominated.
Then, the specific energy can be approximated to be conserved and expressed as
\begin{equation}
\frac{v^{2}}{2} + w - \frac{GM}{r} = \varepsilon_{\rm out},
\label{eqn:E_excretion_flow}
\end{equation}
where $v$ is the flow velocity, $w$ is the specific enthalpy and $\varepsilon_{\rm out}$ is the constant specific energy of the thick excretion flow.
This specific energy is the quantity given in equation (\ref{eqn:epsilon_out}) on the assumption that the specific energy of the inflowing matter to the accretion ring is zero.
The flow is expected to become supersonic and the terminal velocity, $v_{\infty}$, is estimated as
\begin{equation}
v_{\infty} \simeq \sqrt{2GM/R_{0}} = 9 \times 10^{7} \left(\frac{M}{10\; M_{\odot}}\right) \left(\frac{R_{0}}{10^{11.5} \rm{cm}}\right)^{-1} \mbox{ cm s}^{-1}.
\label{eqn:v_infty}
\end{equation}

The accretion ring precesses with the period $T_{\rm p}$ = 162.5 days around the rotational axis of the binary and the half opening angle of the precessing cone, $\theta_{\rm p}$, is 20$^{\circ}$.
Because of the precession of the accretion ring, all the flows: the excretion flows, accretion flows and the jets can be considered to precess simultaneously.

\section{Interpretations of observations}

\subsection{Precession of the accretion ring}
The observations of the jet motions clearly indicate the precession of the jet axis.
Considering that the jets are ejected in the directions perpendicular to the accretion disk, 
Whitemire and Mates (1980), van den Heuvel et al. (1980) and Katz (1980) proposed the slaved disk model for SS433 in which the disk precesses slaved to the precession of the companion star but the stability of the precession is questionable.
Inoue (2012; 2019) discusses that the precession of the accretion ring induced by the tidal force from the companion star is rather stable.

The power spectral analysis of the Doppler shifts observed from the SS433 jets by Katz et al. (1982) shows presence of two periods at 5.8 days and 6.3 days.
Katz et al. (1982) explain the 6.3 day period but fail the 5.8 day period in the frame of the slaved disk model, while Inoue (2012) gives the simpler interpretation of both the two periods by the nodding motion of the jet axis.

\subsection{Hard X-ray emission}
Cherepashchuk et al. (2013) present the hard X-ray (above 20 keV) light curve folded with the precession period.
It looks similar to the X-ray light curve of Her X-1 folded with the super-orbital period of 35 days (see e.g. Inoue 2019).
The super-orbital light curves of Her X-1 and two other X-ray pulsars are shown to be well reproduced by a model in which X-rays from a compact object towards us periodically obscured by a precessing ring at the outermost part of an accretion disk around the central object by Inoue (2019).
These suggest that the periodic obscuration of the central X-ray source by the precessing accretion ring takes place even in SS433.
If so, the hard X-rays should come from an inner region than the accretion ring.
Indeed, the analysis of the shape and width of the primary hard-X-ray eclipses indicates that the hard X-ray emission (20 - 100 keV) is formed in an extended, hot, quiasi-isothermal corona on an accretion disk (Cherepashchuk et al. 2009).

The hard X-ray emission above $\sim$20 keV has the luminosity of several times 10$^{35}$ erg s$^{-1}$ (Cherepashchuk et al. 2003) and the power law spectrum with the photon index $\sim$3 or larger (Cherepashchuk et al. 2009: 2013).
These characteristics are consistent with a picture that the accretion disk in SS433 is in the slim disk configuration (Inoue 2022a) and that the soft X-rays from the disk is completely obscured by the photo-electric absorption, while the hard X-rays are attenuated by electron scattering.
Kubota and Makishima (2004) analyzed the data of the black hole binary XTE J1550-564 during the outburst and showed that the spectrum of the source in the slim disk configuration has a steep hard tail above the main soft component and that the luminosity of the power law component is about 1\% of that of the soft component apparently saturating at the Eddington luminosity.
If we assume the mass of the compact object in SS433 is $\sim$10 $M_{\odot}$, the soft X-ray luminosity should be $\sim 10^{39}$ erg s$^{-1}$ and thus the hard X-ray luminosity is expected to be $\sim 10^{37}$ erg s$^{-1}$.
Hence, the observed hard X-ray luminosity of several times 10$^{35}$ erg s$^{-1}$ requires an intensity attenuation by a factor slightly larger than 10.
A part of the attenuation would be given by the accretion ring and the other part would be by the excretion belt, which is discussed is subsection \ref{Obscuration}.

\subsection{Jet ejection}
Recently, Inoue (2022b) proposes a mechanism for the innermost region of a slim disk to steadily eject the bi-polar jets, and the model can be applied to the SS433 jets.

Most observations indicate that the slim disk intermittently appears in the limit cycle activities in the very high state (see section 5 of Inoue 2022a), while 
the jet activity of SS433 seems to be basically steady.
Thus, if the jets really comes from the innermost region of the slim disk, 
it is required that the slim disk steadily exists in the SS433 system.
Indeed, the activities of type II bursts from the Rapid Burster which are understood to be the result of the limit cycle activity (see e.g. subsection 5.3 in Inoue 2022a) appear in the decay portions of the outbursts and the cyclic period becomes shorter as the source intensity decreases (Guerriero et al. 1999).
These are naturally interpreted by a picture that the recurrent period of the limit cycle becomes so long as for the source to practically be in the steady slim disk phase when the accretion rate is very high.

\subsection{Obscuration of the central engine}\label{Obscuration}
X-ray observations of the jets clearly indicate that the observed soft X-rays below $\sim$10 keV come from the jets and the bright soft X-rays expected from the central part of the accretion disk should be almost completely absorbed via the photoelectric absorption.
Even the X-rays observed from the jets are limited to those from the region with the plasma temperature below $\sim$20 keV (Kotani et al. 1996) and then the central part with the higher temperature should be obscured.
It is further pointed out by Kotani et al. (1996) that the receding jet is more obscured than the preceding jet and that the matter responsible for the obscuration should have a size significantly larger than the accretion disk.

The matter outflowing through the thin excretion disk is likely to significantly contribute to the obscuration of the central engine and the central part of the jets.
As discussed earlier, the matter is considered to eventually form the excretion belt along the Roche lobe and to finally merge with the atmosphere of the optical star (see figure \ref{fig:ExcretionBelt}).
Since the excretion disk precesses with the precessing angle, $\theta_{\rm p} \simeq 20^{\circ}$, the excretion belt would have the width of $\sim 2\; R_{1} \sin \theta_{p}$ and the total flow length of $\sim 2\pi R_{1}$, and thus the total surface area would be $\sim 4\pi R_{1}^{2} \sin \theta_{\rm p}$.
Then, if the staying time of the matter in the excretion belt is assumed to be the Keplerian circular period around the compact star given as $2\pi R_{1}^{3/2}/(GM_{\rm x})^{1/2}$, the average optical depth for the Thompson scattering of the excretion belt, $\tau_{\rm eb}$, is approximately estimated to be as
\begin{equation}
\tau_{\rm eb} \simeq \frac{\dot{M}_{\rm ex,\; d}}{2 (GM_{\rm x})^{1/2} R_{\rm 1}^{1/2} \sin \theta_{\rm p}} \frac{\sigma_{\rm T}}{m_{\rm p}},
\label{eqn:tau_eb}
\end{equation}
where $\sigma_{\rm T}$ is the cross section for the Thomson scattering.
The $\tau_{\rm eb}$ value for $\dot{M}_{\rm ex,\; d} \simeq 10^{20}$ g s$^{-1}$, $M_{\rm x} \simeq 10\; M_{\odot}$ and $R_{1} \simeq 10^{12}$ cm is 1.6, which is consistent with the expectation for the obscuration.

Note that the net photoelectric cross section per hydrogen atom of matter with the cosmic abundance steeply decreases as the photon energy increases and is about the same as the Thomson cross section at $\sim$10 keV (see e.g. Morrison \& McCammon 1983).
Hence, the situation can reasonably be understood that the soft X-rays below $\sim$10 keV from the central engine are completely absorbed, while the hard X-rays above $\sim$10 keV are only attenuated.

\subsection{Interaction between the jet and the disk wind}\label{Jet-DW}
As shown in section \ref{Introduction}, the observed properties of jets around the 10$^{14}$ to 10$^{15}$ cm distance are currently explained by the interaction of the jets with the disk wind.
The radio structures in the plane perpendicular to the jets (Paragi et al. 1999; Blundell et al. 2001) support the presence of the disk wind.
In the frame of the accretion ring concept, the disk wind can naturally be explained by the thick excretion flow.
Hereafter, we call the thick excretion flow as the disk wind following the current terminology.

\begin{figure}
 \begin{center}
  \includegraphics[width=12cm]{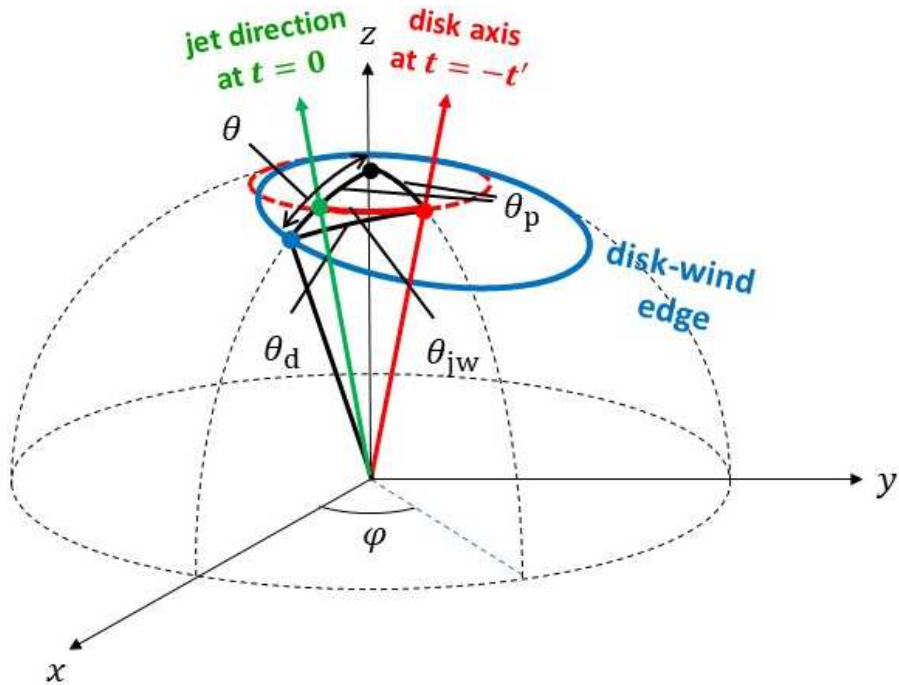}  
 \end{center}
\caption{The isosceles spherical triangle formed by the $z$ axis, the jet direction at $t=0$ which is on the $x$ - $z$ plane and the disk axis at $t=-t'$ which rotates an angle $\varphi$ around the $z$ axis from the $x$ - $z$ plane.  Another spherical triangle formed by the $z$ axis, the disk axis at $t= -t'$ and the crossing point of the small circle depicted by the disk wind edge around the disk axis with the $x$ - $z$ plane is also drawn for the calculations in appendix \ref{DiskWindEdge}.  The inclination angle of the precession axis from the $z$ axis is $\theta_{\rm p}$, the angular radius of the disk edge circle is $\theta_{\rm d}$ and the angle of the crossing point from the $z$ axis is $\theta$.}
\label{fig:DiskWindEdge}
\end{figure}

Let us consider a situation in which an element of the jet is ejected with the velocity, $v_{\rm j}$, at a moment of $t$ = 0 in a radial direction on the surface of the precessing cone with the opening half-angle of $\theta_{\rm p}$ and the rotational axis of the accretion ring yielding the disk wind coincides with the jet direction at $t$ = 0.
The time for the jet element to reach a distance, $r$, from the center, $t_{\rm je,\; r}$, is given as $t_{\rm je,\; r} = r/v_{\rm j}$.
The disk-wind matter around the jet element at $r$ is considered to have started from the center at $t_{\rm dw,\; r,\; 0} = t_{\rm je,\; r} - r/v_{\rm dw} = - [ 1 - (v_{\rm dw}/v_{\rm j})] r/v_{\rm dw}$.
Hence, a difference of the precession phase angle of the disk-wind matter from that at $t=0$, $ \varphi$, is calculated as
\begin{equation}
\frac{\varphi}{2 \pi}  = \frac{t_{\rm dw,\; r,\; 0}}{T_{\rm p}} = -\frac{r/v_{\rm dw}}{T'_{\rm p}},
\label{eqn:DeltaPsi}
\end{equation}
where
\begin{equation}
T'_{\rm p} = \left( 1 - \frac{v_{\rm dw}}{v_{\rm j}} \right)^{-1} T_{\rm p} \simeq T_{\rm p}.
\label{eqn:T'_p}
\end{equation}
Then, the angle, $\theta_{\rm jw}$, between the jet-element direction and the normal axis of the disk wind at $r$ can approximately be estimated from the isosceles spherical triangle drawn in figure \ref{fig:DiskWindEdge} as
\begin{equation}
\cos \theta_{\rm jw} = \cos^{2} \theta_{\rm p} + \sin^{2} \theta_{\rm p} \cos \varphi.
\label{eqn:theta_jw}
\end{equation}
As seen from this equation, $\theta_{\rm jw}$ increases from zero to the maximum of 2$\theta_{\rm p} \simeq 40^{\circ}$ as $\varphi$ increases from zero to $\pi/2$ in association with the increase of $r$ from zero to $r_{\rm opt}$ given as
\begin{equation}
r_{\rm opt} \simeq v_{\rm w} \frac{T_{\rm p}}{2} = 7.0 \times 10^{14} \left(\frac{v_{\rm w}}{10^{8}\mbox{cm s}^{-1}}\right) \mbox{ cm}.
\label{eqn:r_max}
\end{equation}
Since the elevation angle of the matter in the disk wind from the equatorial plane is $(\pi/2) - \theta_{\rm jw}$, the density of the wind matter interacting the jet matter is expected to increase as $\theta_{\rm jw}$ increases.  
This can well explain the optical line activity strengthened at the distance around several times 10$^{14}$ cm as briefly introduced in section \ref{Introduction}.

\begin{figure}
 \begin{center}
  \includegraphics[width=12cm]{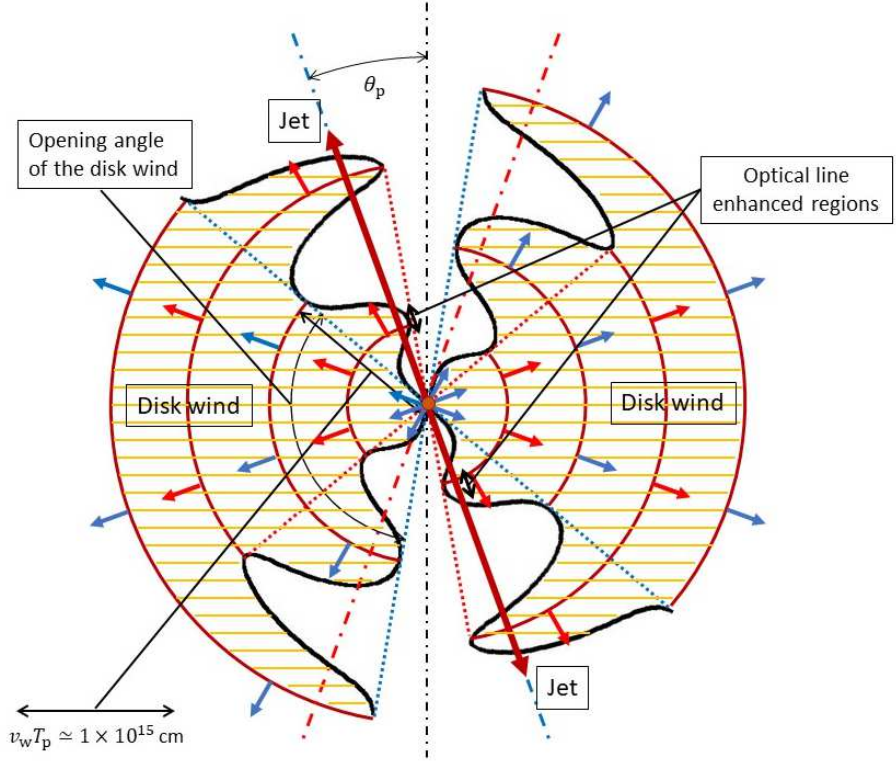}  
 \end{center}
\caption{The cross section of the disk wind on the meridian plane including the angular-momentum axis of the precession and the direction of the jet ejection at a moment.  The opening angle of the disk wind is assumed to be $2 \times 60^{\circ}$.  A bullet of the jet starts from the center, and first interacts with the-disk wind matter around the distance of $v_{\rm w}\; T_{\rm p}/2$, which could cause the enhancement of the optical lines}\label{fig:OpticalZone}
\end{figure}

The above configurations are schematically drawn in figure \ref{fig:OpticalZone}.
This figure exhibits the shape of the disk wind edge on the meridian cross section along the jet axis at a certain moment, where the elevation angle of the disk wind edge is assumed to be 60$^{\circ}$ (the opening angle of the wind $\sim 120^{\circ}$).
The typical scale of this figure is $v_{\rm w} T_{\rm p}$ and the jets run almost instantaneously over this scale since the jet speed is much faster than $v_{\rm w}$.
Hence, this figure approximately shows how the jets intersect with the disk wind.
See appendix \ref{DiskWindEdge} for the calculation of the shape of the disk wind edge.

When the jet matter approaches to the distance of
\begin{equation}
v_{\rm w} T_{\rm p} = 1.4 \times 10^{15} \left(\frac{v_{\rm w}}{10^{8}\mbox{cm s}^{-1}}\right) \mbox{ cm},
\label{eqn:r_min}
\end{equation}
its position becomes perpendicular to the equatorial plane of the disk wind at that position and the ambient density of the disk wind is likely to be very low.
When it further advances to the distance of (3/2) $v_{\rm w} T_{\rm p} \simeq 2.1 \times 10^{15}$ cm, $\theta_{\rm jw}$ again increases to 40$^{\circ}$.
However, the density of the ambient disk wind in the cylinder-like region through which the jet has once passed is expected to be kept very low after sweeping up of the disk wind matter by the jet as discussed in subsection 5.2 of Fabrika (2004).  

The matter in a particular region of the disk wind meets the jet matter every precession period as it advances outward.
It should be shoved away at the first encounter with the jet and an empty cylinder-like space should appear and co-move with the ambient disk-wind matter.
Since the radius of the cylinder-like region is $r\; \Theta_{\rm j}$ at a distance, $r$, the time scale for the ambient matter to refill the empty region, $t_{\rm rf}$, could be approximated as $t_{\rm rf} \sim (r\; \Theta_{\rm j})/c_{\rm s}$, where $c_{\rm s}$ is the sound velocity of the ambient disk-wind matter.
When the refilling time scale is much longer than the precession period, such a situation as the density in the cylinder-like region is so low as for the interaction with the jet to be significantly weak can be considered to be kept at the successive encounters after the first one.

Considering that the cooling time of the disk-wind matter is much shorter than the traveling time to the distance of $\sim 10^{15}$ cm, the temperature of the wind matter should be no higher than 10$^{4}$ K and so the sound velocity, $c_{\rm s}$, should be no larger than 10$^{6}$ cm s$^{-1}$.
Hence, we can say that the swept-up situation is kept in a region of
\begin{equation}
r \gg c_{\rm s}\; T_{\rm p} = 1.4 \times 10^{15} \left(\frac{c_{\rm s}}{10^{6} \mbox{cm s}^{-1}}\right).
\label{eqn:r_swept}
\end{equation}
For $c_{\rm s} \lesssim 10^{6}$ cm s$^{-1}$, the swept-up situation in the disk wind can be expected to be kept after the first encounter with the jet.
This also supports the argument for the radio brightening zone because of the free expansion of the clouds in the jet as in subsection 5.2 of Fabrika (2004).

\subsection{Interaction of the disk wind with the SNR matter}\label{DW-SNR}
It is known that SS433 is in the radio nebula W50 which has the spherical structure with the radius of $\sim$40 pc except the elongated structures in the east and west directions.
This nebula is currently considered to be a SNR.

The disk wind radially flows within a boundary elevation angle from the equatorial plane.
Since the ram pressure at the front  of the disk wind decreases as it advances pushing the SNR matter, the wind front should significantly be decelerated by the SNR matter at some distance.
As discussed in appendix \ref{StationaryShock}, the quasi-steady situation is expected to be realized when the position of the backward shock gets stationary.
The distance of the stationary backward shock, $r_{\rm bs,\; dw}$, is determined from $A$ = 1/9 and $A$ is given in equation (\ref{eqn:Def-A}).
Here, $\rho_{\rm st}$ in appendix should be replaced to the density of the SNR matter in W50, $\rho_{\rm snr}$, and $\rho_{\rm sf,\; bs}$ should be the density of the disk wind on the upstream side of the backward shock, $\rho_{\rm dw,\; bs}$, which is given as
\begin{equation}
\rho_{\rm dw,\; bs} \simeq \frac{\dot{M}_{\rm dw}}{4\pi r_{\rm bs,\; dw}^{2} \cos \alpha_{\rm dw}  v_{\rm w}},
\label{eqn:rho_dw,bs}
\end{equation}
where $\dot{M}_{\rm dw} (= \dot{M}_{\rm ex,\; f}$) is the mass flow rate through the disk wind, and $\alpha_{\rm dw}$ is the elevation angle of the upper boundary of the disk wind from the equatorial plane.
$\xi$ is set unity for the disk wind.
From these equations, the position of the backward shock, $r_{\rm dw,\; bs}$, is approximately calculated as
\begin{equation}
r_{\rm dw,\; bs} \simeq 6.6 \times 10^{17} \left(\frac{\dot{M}_{\rm dw}}{10^{19}\mbox{ g s}^{-1}}\right)^{1/2} \left(\frac{v_{\rm w}}{10^{8} \mbox{ cm s}^{-1}}\right)^{-1/2} \left(\frac{\rho_{\rm snr}}{10^{-24} \mbox{ g cm}^{-3}}\right)^{-1/2} \mbox{ cm},
\label{eqn:rho_dw,bs_Cal}
\end{equation}
where $\alpha_{\rm dw} = 60^{\circ}$ is adopted.
Hence, the stationary backward shock is expected to appear at the distance of $7 \times 10^{17}$ cm for $\dot{M}_{\rm dw} \sim 10^{19}$ g s$^{-1}$, $v_{\rm w} \sim 10^{8}$ cm s$^{-1}$ and $\rho_{\rm snr} \sim 10^{-24}$ g cm$^{-3}$.

The heating of the disk wind matter should take place behind the backward shock and could induce refilling the the matter-swept-up region in the jet-path during a cycle of the jet precession, as discussed in the next paragraph.
This could cause reheating the jet matter through the interaction with the refilled disk mater, possibly originating the jet structures observed in the X-ray band around $\sim 10^{17}$ cm distance (Migliari et al. 2002).
This distance of $\sim 10^{17}$ cm seems smaller than the above estimated distance of the backward shock for the reasonable parameters.
However, since the interaction should takes place in a region with the elevation angle above 50$^{\circ}$ from the disk-wind equatorial plane, it is expected that the ram pressure of the disk wind is significantly weaker there than the typical value and that the front shock distance is significantly smaller than the typical value.

As discussed in the previous subsection, the time scale for the ambient disk-wind matter to refill the empty region along the jet path at a distance, $r$, is approximated as $r\; \Theta_{\rm j}/c_{\rm s}$.
Then, the refilling could be effective when the refilling time scale is no longer than the precession period, and we can get the following condition for the effective refilling to happen as
\begin{equation}
r \lesssim \frac{c_{\rm s} T_{\rm p}}{\Theta_{\rm j}} = 7 \times 10^{16} \left(\frac{c_{\rm s}}{10^{8} \mbox{ cm s}^{-1}}\right) \mbox{ cm}.
\label{eqn:r-rh}
\end{equation}
Considering $c_{\rm s} \sim v_{\rm dw} \sim 10^{8}$ cm s$^{-1}$ on the subsonic side of the backward shock, this is roughly consistent with the reheated position to emit X-rays within the accuracy of the present estimation.

\begin{figure}
 \begin{center}
  \includegraphics[width=12cm]{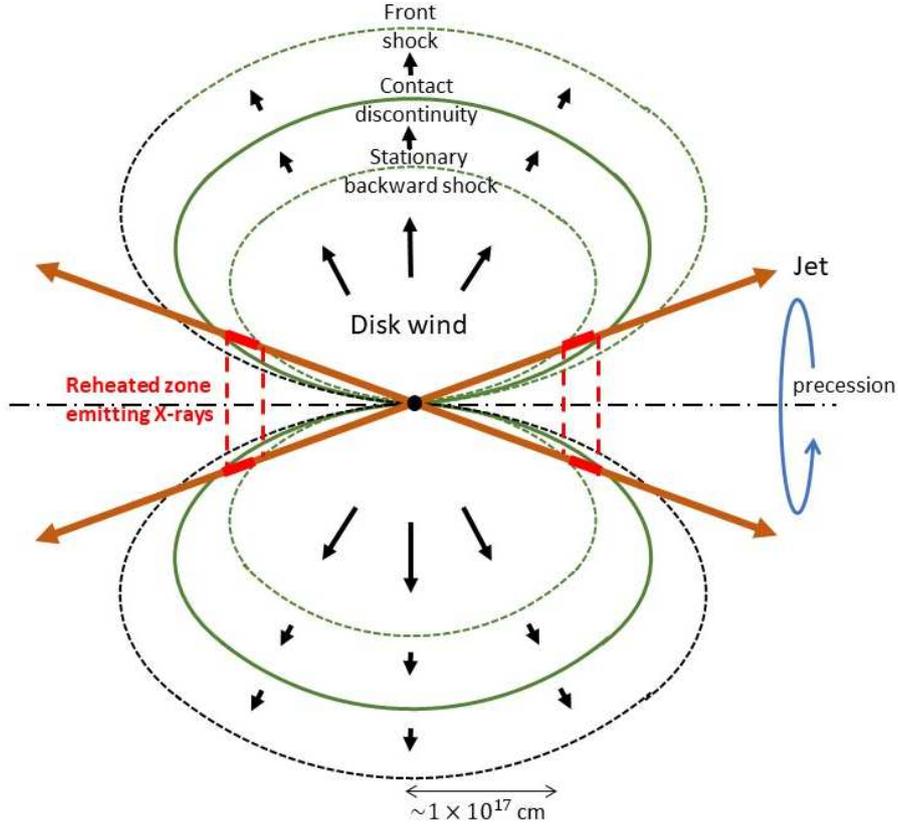}  
 \end{center}
\caption{The schematic cross section of the interaction between the disk wind and the SNR matter, and the position of the reheated zone emitting X-rays as the result of the interaction of the jet with the shocked disk-wind matter.}
\label{fig:X-rayReheatingZone}
\end{figure}

The schematic view of the interaction between the disk wind and the SNR matter and that between the jet and the shocked disk wind is shown in figure \ref{fig:X-rayReheatingZone}.

\subsection{Interaction of the jets with the SNR matter}
Similarly to the case of the interaction of the disk wind with the SNR matter, 
the front of the jet should significantly be decelerated by the SNR matter at some distance.
According to appendix \ref{StationaryShock} again, the distance of the stationary backward shock in the jet, $r_{\rm bs,\; j}$, is determined from $A$ = 1/9 and $A$ is given in equation (\ref{eqn:Def-A}).
In this case, $\rho_{\rm st}$ should be the density of the SNR matter, $\rho_{\rm snr}$, again but $\rho_{\rm sf,\; bs}$ should be the density of the jet on the upstream side of the backward shock, $\rho_{\rm j,\; bs}$, which is given as
\begin{equation}
\rho_{\rm j,\; bs} \simeq \frac{\dot{M}_{\rm j}}{2\pi r_{\rm bs,\; dw}^{2} \Theta_{rm j}^{2}  v_{\rm j}},
\label{eqn:rho_j,bs}
\end{equation}
where $\dot{M}_{\rm j}$ is the total mass flow rate through the two opposite jets, and $\Theta_{\rm j}$ is the half opening angle of the jet.
$\xi$ should be significantly smaller than unity here.
From these equations, the position of the backward shock, $r_{\rm j,\; bs}$, is approximately calculated as
\begin{equation}
r_{\rm j,\; bs} \simeq 11 \left(\frac{\xi}{0.1}\right)^{-1/2} \left(\frac{\dot{M}_{\rm j}}{10^{20}\mbox{ g s}^{-1}}\right)^{1/2} \left(\frac{\Theta_{\rm j}}{1.2^{\circ}}\right)^{-1/2} \left(\frac{\rho_{\rm snr}}{10^{-24} \mbox{ g cm}^{-3}}\right)^{-1/2}\mbox{ pc}.
\label{eqn:rho_j,bs_Cal}
\end{equation}
Hence, the stationary backward shock is expected to appear at the distance of $\sim$ 10 pc for $\xi \sim 0.1$, $\dot{M}_{\rm j} \sim 10^{20}$ g s$^{-1}$, $\Theta_{\rm j} \simeq 1.2^{\circ}$ and $\rho_{\rm snr} \sim 10^{-24}$ g cm$^{-3}$.

If the position of the backward shock in the jet is located around 10 pc, most of the momentum carried by the jet is likely to be transfered to the SNR matter which is shoved by the front of the jet, as the front advances further in the SNR matter with the velocity of $V_{\rm cd} \simeq v_{\rm j}/4 \simeq 2 \times 10^{9}$ cm s$^{-1}$.
The SNR matter pushed into the precessing cone of the jet is expected to gather along the cone axis and to advance toward the interstellar space with the speed of $\sim 10^{9}$ cm s$^{-1}$.
This could be the origin of the elongated structures seen in radio (Dubner et al. 1998) and X-rays (Watson et al. 1983) in the east and west directions from the main structure in W50.

\begin{figure}
 \begin{center}
  \includegraphics[width=14cm]{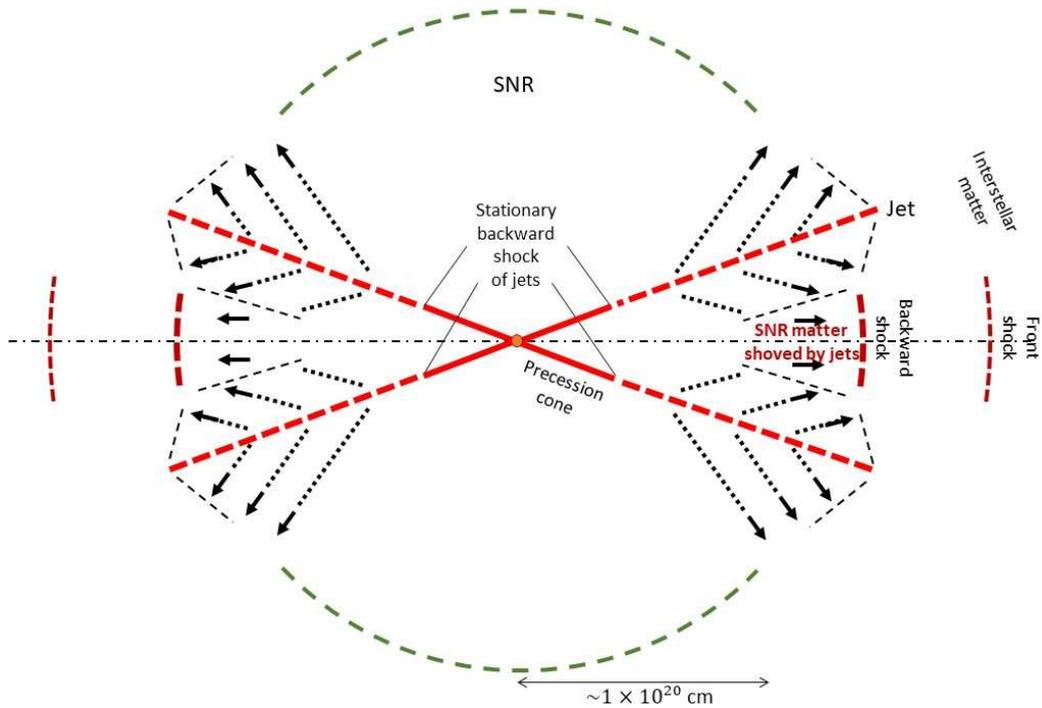}  
 \end{center}
\caption{The schematic cross section of the interaction between the jets and the SNR matter.}
\label{fig:ElongatedStructures}
\end{figure}

The above arguments are schematically visualized in figure \ref{fig:ElongatedStructures}.

\section{Summary and discussion}
\begin{figure}
 \begin{center}
  \includegraphics[width=16cm]{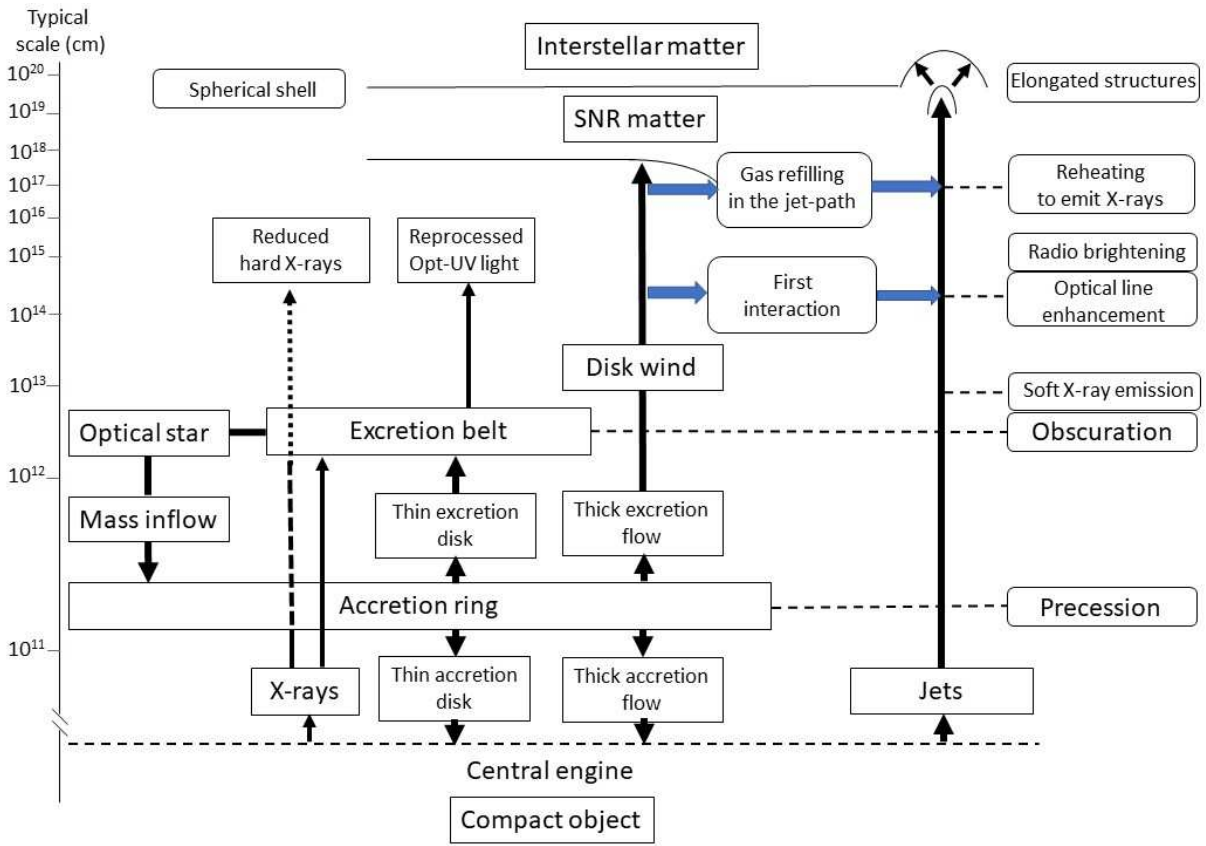}  
 \end{center}
\caption{The block diagram of elements and events in the SS433 -- W50 system.}\label{fig:BlockDiagram}
\end{figure}

The present study is summarized in figure \ref{fig:BlockDiagram} as a block diagram of the elements discussed above.

The present study starts from adoption of the ``accretion ring" concept introduced by Inoue (2021a) to the SS433 system.
It predicts that the accretion ring bears a thick excretion flow and a thin excretion disk.
The thick excretion flow is considered to eventually becomes a supersonic flow with the terminal velocity of 10$^{8}$ cm s$^{-1}$, and plays an important role as the disk wind in the present scenario.
The thin excretion disk tends to form a circular ring with the radius four times as large as that of the accretion ring if there is no gravitational effect from the normal star. 
In the present scenario for the SS433 system, the circular ring appears in the form of the excretion belt, which gives a significant contribution to the obscuration of the central engine.

The accretion ring is also discussed to be a place where a precessing motion is naturally excited (Inoue 2012; 2019).
The precession seen in the jet motion of SS433 can be considered to originate at the accretion ring.
Indeed, the hard X-ray light curve folded with the precession period resembles with the X-ray light curve of Her X-1 folded with the super-orbital period of $\sim$35 days, while Inoue (2019) reveals that the X-ray light curve of Her X-1 is well reproduced by a model in which X-rays from a compact object towards us are periodically obscured by a precessing accretion ring around a central object.
Since the elevation angles of the line of sight to the binary plane are as small as about 10$^{\circ}$ or lower both for SS433 and Her X-1,the resemblance in the light curve between SS433 and Her X-1 suggests that the hard X-rays on the line of sight from SS433 are periodically obscured by the precessing accretion ring. 

In spite of such similarities between SS433 and Her X-1 as mentioned above, distinct differences of SS433 from Her X-1 are the presence of the jets and the obscuration of the central engine.
These are likely to be due to the different situation of the accretion rate:
The supercritical accretion of $\sim 10^{20}$ g s$^{-1}$ is considered to happen on to the compact object with several solar mass in SS433, while the accretion rate in Her X-1 is no larger than 10$^{18}$ g s$^{-1}$ against the neutron star of $\sim 1$ solar mass.
Such a high accretion rate is expected to cause the jet ejection from the central part of the slim disk (Inoue 2022b).
The estimation of the optical depth for Thomson scattering of the excretion belt in equation (\ref{eqn:tau_eb}) indicates $\tau_{\rm eb} > 1$ for $\dot{M} \sim 10^{20}$ g s$^{-1}$ in the case of SS433, while $\tau_{\rm eb}$ is estimated to be 0.09 for $\dot{M} \sim 10^{18}$, $M_{\rm x} \sim 1\; M_{\odot}$ and $R_{1} \sim 3 \times 10^{11}$ cm in the case of Her X-1.

The jets begin to be observed in soft X-rays after they go out of the central obscured region, and the diagnostics of the X-ray lines are consistent with the adiabatically and radiatively cooling model of the conical jets (Kotani et al. 1996).

As the jet goes out from $\sim 10^{13}$ cm to $\sim 10^{14}$ cm, the temperature decreases and the mainly observed wave-band shifts from the X-ray band to the optical band.
The time variabilities of the optical lines on a time scale of days begins $\sim 10^{14}$ cm, have the activity maximum at several times 10$^{14}$ cm and ends around 10$^{15}$ cm.
These behaviors are consistent with being due to the first interaction of the jet with the disk wind as discussed in subsection \ref{Jet-DW}.

The matter in the certain region in the disk wind which have first encountered with the jet at several times 10$^{14}$ cm should have been shoved away from the jet-path by the jet matter, and the region is likely to get almost empty.
The empty region should co-move with the ambient matter in the disk wind and repeatedly encounter with the jet every precession period after the first encounter.
The time scale for the ambient matter to refill the empty region is, however, estimated to be longer than the precession period in the region given in equation (\ref{eqn:r_swept}), and thus the empty situation of the jet path in the disk wind is kept until the disk wind reaches the distance of $\sim 10^{17}$ cm.

At the $\sim 10^{17}$ cm distance, the jet-path in the disk wind is expected to enter the subsonic region behind the backward shock induced by the interaction with the SNR matter and to be refilled with the ambient disk-wind matter.
The interaction of the jet element with the refilled disk-wind matter could cause reheating the jet element to emit X-rays, as discussed in subsection \ref{DW-SNR}.

The jet-element gets into the SNR region after passing through the backward shock region and the jet-path in the SNR matter is likely to be swept up again.
The ram pressure of the jet decreases as it goes further and is finally braked by the SNR matter, forming the stationary backward shock at the distance of $\sim$10 pc.
The momentum carried by the jet is expected to transfer to the SNR matter in front of the jet.
About a half of the SNR matter shoved by the jet could be collimated to the outward direction along the central axis of the precession cone and advance into the interstellar space, forming the front shock and the backward shock.
This movement could originate the elongated structure on each of the east and west side.

As summarized above, the scenario in this paper can basically explain the overall features of the SS433 -- W50 system.
However, the arguments are yet semi-quantitative.
Further study is obviously needed.





\appendix 
\section{Shape of the disk wind edge}\label{DiskWindEdge}
Let us introduce a three dimensional orthogonal coordinate system ($x$, $y$, $z$) and consider a situation in which the rotationally symmetric axis of the mass ejection (the jets and the disk wind) system precesses along a small circle with a angular radius, $\theta_{\rm p}$ around the $z$ axis, as shown in figure \ref{fig:DiskWindEdge}.
The rotationally symmetric axis is called the disk axis here.

We define $t$ as a time from a moment when the disk axis is right on the $x$ - $z$ plane, and calculate the shape of the disk wind edge on the $x$ - $z$ plane at the moment of $t$ = 0.

The disk axis of the disk wind ejected at $t = -\; t'$ rotates backward by an angle $\varphi $ from the $x$ - $z$ plane written as
\begin{equation}
\varphi = \frac{2\pi}{T_{\rm p}} \ t'.
\label{eqn:phi-Def}
\end{equation}
If the zenith angle of the disk wind edge from the disk axis is $\theta_{\rm d}$, the edge of the disk wind ejected $t'$ ago should form the small circle with the angular radius of $\theta_{\rm d}$ around the disk axis at that time.
Then, the angle of the direction of the disk wind edge on the $x$ - $z$ plane at $t$ = 0 from the $z$ axis, $\theta$, is calculated with the help of the laws of sine and cosine for the non-isosceles spherical triangle as shown in figure \ref{fig:DiskWindEdge} as
\begin{equation}
\cos \theta = \frac{\cos \theta_{\rm p} \cos \theta_{\rm d} + \sin \theta_{\rm p} \sin \theta_{\rm d} \cos \varphi \sqrt{1 - \sin^{2} \theta_{\rm p} \sin^{2} \varphi/\sin^{2} \theta_{\rm d}}}{1-\sin^{2} \theta_{\rm p} \sin^{2} \varphi}.
\label{eqn:cos-theta}
\end{equation} 

On the other hand, the matter in the disk wind ejected $t'$ ago is considered to advance radially over the distance, $r$, by $t$ = 0, given as 
\begin{equation}
r = v_{\rm dw} t',
\label{eqn:r}
\end{equation}

As the results, the position of the edge of the disk wind ejected $t'$ ago on the $x$ - $z$ plane at the moment, $x(t')$ and $z(t')$ are calculated as
\begin{eqnarray}
x(t') &=& r \sin \theta, \\
z(t') &=& r \cos \theta.
\end{eqnarray}
Figure \ref{fig:OpticalZone} plots the calculated results of ($x$, $z$) for $\theta_{\rm p} = 20^{\circ}$ and $\theta_{\rm d}= 30^{\circ}$ over two precession periods for $t'$.

\section{Interaction of supersonic flow with the ambient matter}\label{StationaryShock}
First, we consider the interaction of a highly supersonic flow with stationary matter in front of the flow, one-dimensionally.
This is the case in which the disk wind collides with the ambient SNR matter.

Let $\rho_{\rm st}$, $\rho_{\rm sf}$, $v_{\rm sf}$ be the density of the stationary matter, the density and velocity of the supersonic flow. 

A front shock, a contact discontinuity and a backward shock should appear and 
the velocities of them are designated as, $V_{\rm fs}$, $V_{\rm cd}$ and $V_{\rm bs}$ respectively.
The strong shock condition is applied to the shocks and the specific heat ratio is set to be 5/3.
It is also assumed that the regions behind the front shock and the backward shock have the uniform pressure, $P_{\rm cd}$, and the uniform velocity of $V_{\rm cd}$, sandwiching the contact discontinuity surface.

Then, from the equations for the mass continuity and the momentum continuity across the shock surface, we get 
\begin{equation}
V_{\rm fs} = \frac{4}{3} V_{\rm cd},
\label{eqn:V_fs-V_cd}
\end{equation}
and
\begin{equation}
P_{\rm cd} = \frac{4}{9} \rho_{\rm st} V_{\rm cd}^{2},
\label{eqn:P_cd_fs}
\end{equation}
for the front shock, and
\begin{equation}
V_{\rm bs} = \frac{4}{3} V_{\rm cd} - \frac{1}{3} v_{\rm sf},
\label{eqn:V_bs-V_cd}
\end{equation}
and
\begin{equation}
P_{\rm cd} = \frac{4}{3} \rho_{\rm sf,\; bs} (v_{\rm sf}-V_{\rm cd})^{2},
\label{eqn:P_cd_bs}
\end{equation}
for the backward shock.
Here, $\rho_{\rm sf,\; bs}$ is the density of the supersonic flow on the upstream side to the backward shock.
Equating equations (\ref{eqn:P_cd_fs}) and (\ref{eqn:P_cd_bs}), we obtain 
\begin{equation}
\frac{4}{9} \rho_{\rm st} V_{\rm cd}^{2} = \frac{4}{3} \rho_{\rm sf} (v_{\rm sf}-V_{\rm cd})^{2}.
\label{eqn:P_cd}
\end{equation}
From equations (\ref{eqn:V_fs-V_cd}), (\ref{eqn:V_bs-V_cd}) and (\ref{eqn:P_cd}), $V_{\rm fs}$, $V_{\rm cd}$ and $V_{\rm bs}$ can be determined for given $\rho_{\rm st}$, $\rho_{\rm sf,\; bs}$ and $v_{\rm sf}$.

Next, we consider a case in which the jet plunges into the SNR matter.
In this case, differently from the first case, the stationary matter should not accumulate in front of the jet but flow out to the side of the jet in the jet rest frame.
As the result, the pressure, $P_{\rm cd}$, loaded to the jet matter by the stationary matter should significantly be reduced from the value in equation (\ref{eqn:P_cd_fs}), and can be expressed as
\begin{equation}
P_{\rm cd} = \xi \frac{4}{9} \rho_{\rm st} V_{\rm cd}^{2},
\label{eqn:P_cd_fs_2}
\end{equation}
where $\xi$ is the reduction factor.
Simultaneously, $V_{\rm fs}$ is considered to roughly equal to $V_{\rm cd}$.
Equations (\ref{eqn:V_fs-V_cd}) and (\ref{eqn:P_cd_bs}) hold even in this case.

Now, equation (\ref{eqn:P_cd}) can be generalized both for the above two cases as
\begin{equation}
\xi \frac{4}{9} \rho_{\rm st} V_{\rm cd}^{2} = \frac{4}{3} \rho_{\rm sf,\; bs} (v_{\rm sf}-V_{\rm cd})^{2},
\label{eqn:P_cd_gen}
\end{equation}
where $\xi$ is 1 in the first case and significantly smaller the unity in the second case.
This equation can be rewritten as
\begin{equation}
x^{2} - \frac{2A}{A-1}x+\frac{A}{A-1}=0.
\label{eqn:QuadraEq_x}
\end{equation}
where
\begin{equation}
x=\frac{V_{\rm cd}}{v_{\rm sf}},
\label{eqn:Def-x}
\end{equation}
and
\begin{equation}
A = \frac{3 \rho_{\rm sf,\; bs}}{\xi \rho_{\rm st}}.
\label{eqn:Def-A}
\end{equation}
The solution of this quadratic equation for $x$ to smoothly pass through the point of $x = 1/2$ (the solution for $A$ = 1) is
\begin{equation}
x = \frac{\sqrt{A}}{\sqrt{A}+1}.
\label{eqn:Sol-x}
\end{equation}

Since the supersonic flow is the jet or the disk wind here, 
the density of the supersonic flow at a distance, $r$, from the center can be approximated to be proportional to $r^{-2}$.
On the other hand, the stationary matter is regarded to be the SNR matter here and the density can be set to be constant independently of the place.
Hence, we can consider that $A \propto r^{-2}$.
When $r$ is sufficiently small, $A$ is much larger than unity and then $x \simeq 1$, i.e. $V_{\rm cd} \simeq v_{\rm sf}$.
However, if the front of the supersonic flow advances so largely as for $A$ to get as small as or smaller than unity, $x$ comes to decrease from unity and the front of the sonic flow is decelerated.
Finally, the quasi-steady state keeping $A$ constant is likely to be realized when the position of the backward shock stops, i.e. $V_{\rm bs} = 0$.
Then, from equation (\ref{eqn:V_fs-V_cd}) and $V_{\rm bs} = 0$, we get $V_{\rm cd} = v_{\rm sf}/4$, i.e. $x$ = 1/4.
We see from equation (\ref{eqn:Sol-x}) that $A$ = 1/9 when $x$ = 1/4.



\end{document}